# A High-Precision Frequency Locking Method Based on All-Phase FFT Demonstrated on a Crystal Oscillator with Rubidium Clock Reference

Qibin Zheng, Kang Xu, Jiacheng Yang, Liguo Zhou, Li Ding, Xianfeng Jiang, Zhaohui Bu

***Abstract*—This article proposes a novel frequency-locking method based on frequency-domain unbiased phase estimation (FDUPE) for high-precision locking. By performing weighted recombination of the acquired data followed by the Fourier transform processing, the phase at the center of the data segment can be estimated without bias, making the method suitable for frequency-locking applications. In this article, the principle of the proposed method is analyzed, and an electronic prototype is developed to experimentally validate its feasibility by utilizing analog-to-digital converters (ADCs) for signal digitization and a field-programmable gate array (FPGA) to implement the FDUPE algorithm. A digital proportional – integral – derivative (PID) controller is also implemented on the FPGA to provide feedback for accurate frequency locking. In the experiment, a 10-MHz voltage-controlled oscillator (VCO) with a free-running Allan deviation of $1 \times 10^{-9}$ at 1 s is used as the device under test (DUT), while a rubidium atomic clock with Allan deviation of $2 \times 10^{-11}$ at 1 s serves as the high-stability reference source. Experimental results show that the proposed system achieves excellent locking performance, reducing the standard deviation of frequency fluctuations from 12.75 mHz root-mean-square (rms) in the free-running state to 0.88 μHz rms after locking. Correspondingly, the Allan deviation at 10 s is reduced from $9.6 \times 10^{-10}$ to $1.45 \times 10^{-14}$, representing a five-order-of-magnitude improvement in frequency stability.**

## I. Introduction

HIGH-stability frequency sources play a crucial role in advanced applications across various fields, including precision metrology [1], [2], optical frequency combs [3], [4], satellite timing and time transfer [5], [6], and communication systems [7], [8]. However, under free-running conditions, frequency sources are inevitably affected by environmental perturbations such as temperature, humidity, and pressure fluctuations [9], [10], which lead to frequency drift and jitter, thereby degrading the accuracy and stability of the overall system. Although stabilization techniques based on ultra-stable optical cavities can provide excellent frequency stability [11], [12], such systems typically involve bulky structures, high costs, and stringent environmental requirements [13], limiting their applicability in many practical scenarios. Furthermore, many applications operate in outdoor environments or on space-constrained platforms, such as satellites, where environmental control is limited, and the system size must satisfy strict constraints while maintaining sufficient frequency stability [14]. Therefore, compact, robust, and high-precision frequency-locking systems are desirable for practical engineering applications.

Miniaturized atomic clocks are commonly employed in compact frequency-locking systems as stable reference sources to discipline the device under test (DUT) [15], [16], such as an oscillator or other frequency source. Early compact frequency-locking systems were mainly implemented using analog circuitry, where frequency-error measurements were typically performed in the time domain. In-phase and quadrature (IQ) demodulation is a representative time-domain approach [17]. However, analog IQ-based locking systems often suffer from environmental sensitivity and limited locking precision [18]. To address these limitations, frequency-locking techniques have gradually evolved toward digital implementations based on microcontroller units (MCUs) and FPGAs [19], [20]. Compared with MCUs, which typically support kilohertz-level control bandwidths [21], field-programmable gate arrays (FPGAs) provide higher bandwidth and greater implementation flexibility, making them more suitable for high-precision and environmentally robust locking applications [22], [23].

Among these digital implementations, digital quadrature demodulation (DQD) has been widely adopted due to its simple structure and ease of implementation [24], [25]. However, its measurement accuracy is limited by the stability and phase noise of the numerically controlled oscillator (NCO), which constrains the achievable locking precision. To mitigate this limitation, our previous study developed an ADC-based dual-mixer time-difference (ADC-based-DMTD) architecture as a compact and fully digital solution for high-resolution frequency-error extraction and frequency locking [26]. By employing synchronous sampling together with a dual-channel differential structure sharing a common NCO, this architecture

Manuscript received xxx xx, 2026; revised xxx xx, xxx2026. This project is supported in part by grants from the National Natural Science Foundation of China (No. 12575204 and 62201464), the Key Science & Technology Project of Anhui Province (No. 202523t06050005) and the National Key R&D Program of China (No. 2023YFF0719200). (Corresponding author: Qibin Zheng and Liguo Zhou).

Qibin Zheng, Kang Xu, Jiacheng Yang, Liguo Zhou, Li Ding, Zhaohui Bu are with the Quantum Medical Sensing Laboratory and School of Health Science and Engineering, University of Shanghai for Science and Technology, Shanghai 200093, China (e-mail: qbzheng@usst.edu.cn, zlg@usst.edu.cn).

Xianfeng Jiang is with the Shanghai Astronomical Observatory, Chinese Academy of Sciences, Shanghai 200030, China.

effectively suppresses common-mode noise and improves measurement resolution. Nevertheless, its time-domain estimation mechanism still limits further improvement in locking precision. Since frequency locking ultimately depends on accurate phase estimation, frequency-domain signal processing provides a promising route to improve frequency-error extraction accuracy in digital frequency-locking systems [27], [28].

However, conventional fast Fourier transform (FFT)-based phase estimation is susceptible to spectral leakage and window-induced phase distortion under finite-length data records, thereby limiting its accuracy in high-precision frequency-locking applications [29], [30]. Although interpolation-based spectral estimation methods can improve parameter estimation accuracy to some extent, they usually introduce additional spectral peak correction, interpolation, or refinement procedures, increasing implementation complexity [31], [32]. By contrast, all-phase fast Fourier transform (APFFT) suppresses spectral leakage and mitigates phase distortion through all-phase preprocessing of the sampled data. By performing weighted recombination of the acquired data prior to the Fourier transform processing, APFFT enables unbiased estimation of the center-sample phase of the data segment, thereby providing a reliable basis for high-precision frequency-error extraction in digital frequency-locking systems [33], [34], [35].

On this basis, this study proposes an APFFT-based frequency-locking method and establishes a frequency-domain unbiased phase estimation (FDUPE) framework for digital closed-loop locking. In the proposed method, APFFT is employed to simultaneously extract the phases of the reference (REF) and DUT signals. The corresponding frequency deviation is then derived from their phase evolution over consecutive measurement intervals and fed into a digital feedback loop to discipline the DUT to the REF signal in real time. To validate the proposed method, a frequency-locking system was implemented on an FPGA platform using a 10-MHz crystal oscillator as the DUT and a rubidium atomic clock as the reference. Experimental results show that the proposed method achieves an Allan deviation of $1.45 \times 10^{-14}$ at an averaging time of 10 s, corresponding to an improvement of approximately five orders of magnitude over the free-running DUT. Compared with the previous time-domain frequency-locking method [26], the proposed approach achieves improved frequency-locking stability under identical experimental conditions. These results demonstrate that the proposed APFFT-based frequency-locking method is an effective and compact solution for high-precision frequency stabilization in practical engineering applications.

This paper is organized as follows. Section II details the architecture of the proposed frequency-locking system and provides a theoretical analysis of the APFFT-based frequency-locking method. Section III presents the simulation results and evaluates the impact of key parameters on measurement precision. Section IV describes the hardware implementation and discusses the experimental results. Finally, Section V concludes the paper and discusses future improvements.

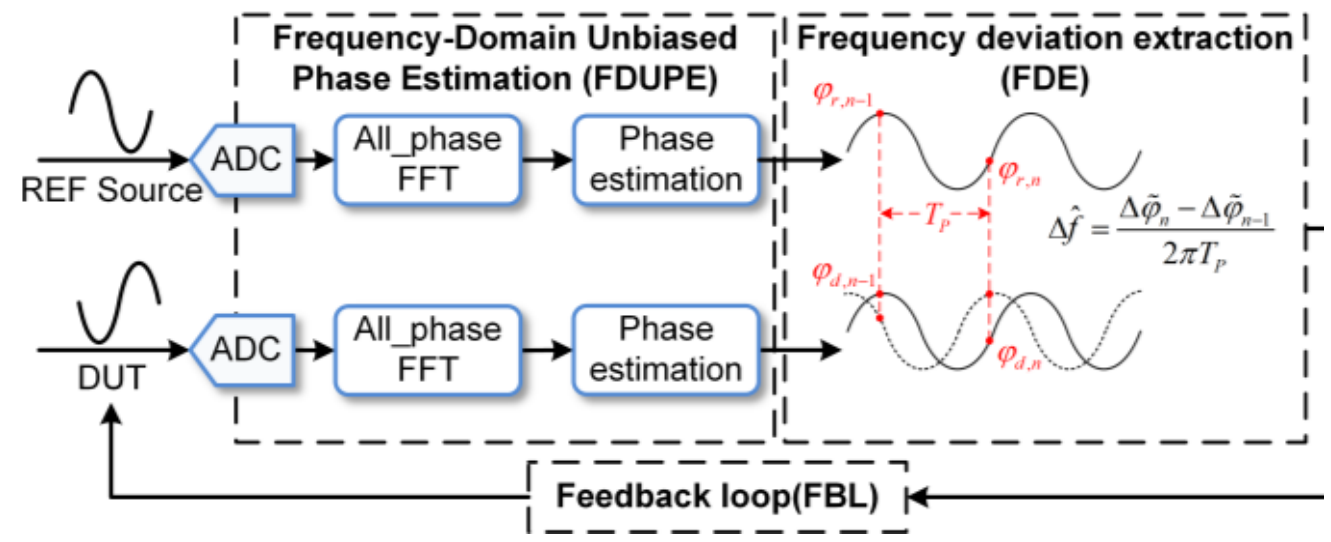

Fig. 1. Principle of the frequency-locking method based on the APFFT.

## II. Methodology

### A. Frequency-Locking Method Based on APFFT

The principle of the proposed frequency-locking method is illustrated in Fig. 1. The method mainly comprises an FDUPE module and a frequency-deviation extraction (FDE) module. Using APFFT, the phases of the REF and DUT signals are estimated with high precision, and the corresponding frequency deviation is then derived. The DUT is disciplined and locked to the reference signal by continuously extracting and feeding back the frequency deviation between the REF and DUT.

Within the FDUPE module, the REF and DUT signals are synchronously digitized by two ADC channels and processed through two parallel APFFT paths. During all-phase preprocessing, a set of overlapping subsequences sharing the same center sample is constructed, center-aligned, and weighted averaged before the FFT operation. This procedure mitigates the phase bias caused by data truncation and boundary effects, so that the phase of the dominant spectral component more accurately represents the signal phase at the center of the observation interval. The resulting phase sequences are then delivered to the FDE module, where the differential phase evolution between the REF and DUT signals over consecutive measurement intervals is converted into the corresponding frequency deviation and fed back to discipline the DUT for frequency locking. Since this frequency-deviation estimation directly relies on the extracted phase sequences, accurate and unbiased phase extraction is essential. The principle of APFFT-based unbiased phase estimation is detailed below.

As shown in Fig. 2, a total of $(2N-1)$ samples, as expressed in (1), are first digitized for APFFT-based phase estimation. Unlike the conventional FFT, which directly applies the Fourier transform to the acquired data segment, APFFT introduces an all-phase preprocessing stage before the FFT operation [33]. Specifically, the $(2N-1)$ samples are divided into a set of overlapping $N$-point subsequences according to the segmentation rule in (2), with all subsequences sharing the same center sample. Through center alignment with respect to the common center sample and weighted averaging according to (4), the overlapping subsequences are reorganized into an $N$-point all-phase sequence. An $N$-point FFT is then performed on the reconstructed sequence to extract the spectral phase. This procedure mitigates the phase bias caused by data truncation and boundary effects, thereby enabling unbiased phase estimation at the center of the observation interval.

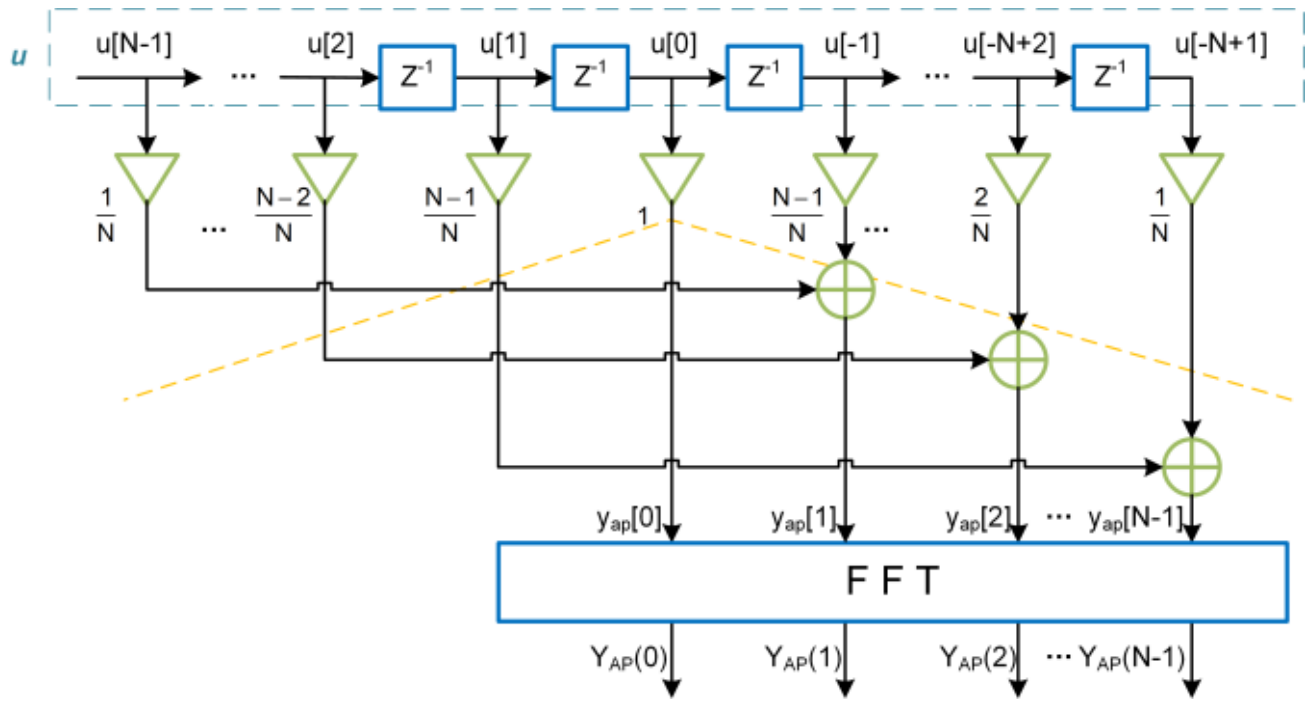

Fig. 2. Block diagram of the $N$-point APFFT algorithm.

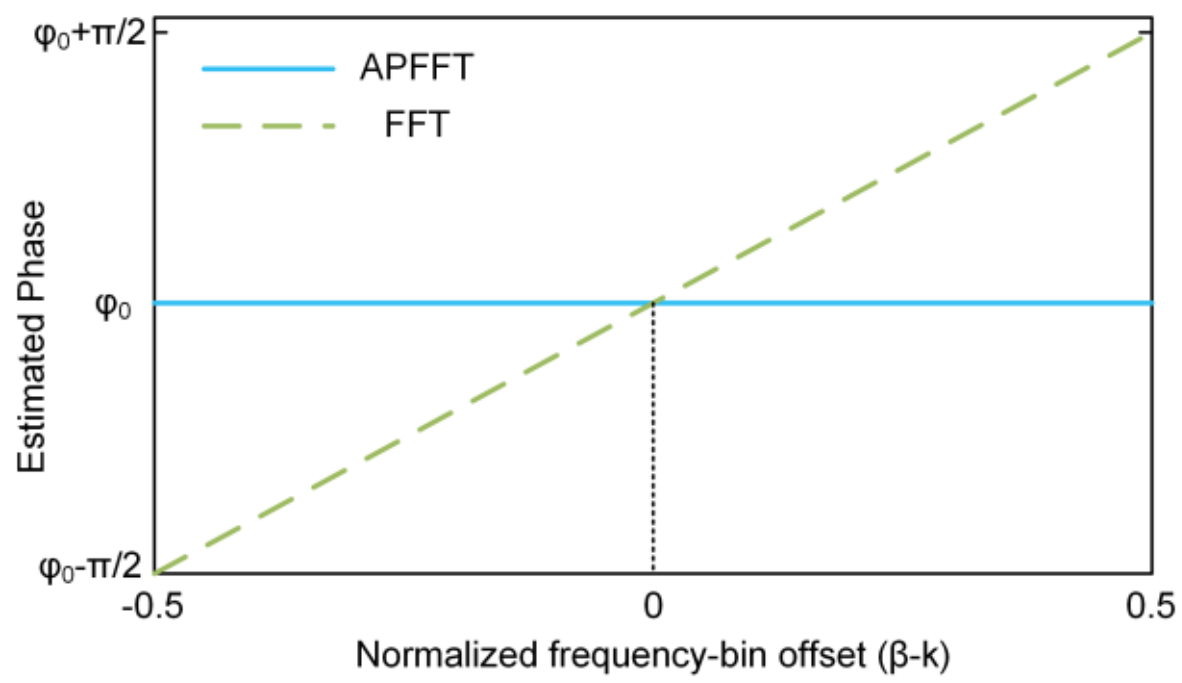

Fig. 3. Comparison of FFT- and APFFT-based phase estimates versus the normalized frequency-bin offset ($\beta-k$), where the phase is evaluated at the center sample $u[0]$.

A (2$N$−1)-sample data segment is expressed as

$$\boldsymbol{u} = \{u[-N+1],...,u[-1],u[0],u[1],...,u[N-1]\}^T. \tag{1}$$

Based on (1), the (2$N$−1) samples are divided into a set of $N$ overlapping $N$-point subsequences, which are written as

$$\begin{aligned}\boldsymbol{u}_0 &= \{u[0],u[1],...,u[N-1]\}^T\\ \boldsymbol{u}_1 &= \{u[-1],u[0],...,u[N-2]\}^T\\ &\vdots\\ \boldsymbol{u}_{N-1} &= \{u[-N+1],u[-N+2],...,u[0]\}^T.\end{aligned} \tag{2}$$

Since all subsequences contain the center sample $u[0]$, each subsequence is circularly shifted to align $u[0]$ with the first position. The resulting aligned subsequences are expressed as

$$\begin{aligned}\boldsymbol{u}'_0 &= \{u[0],u[1],...,u[N-1]\}^T\\ \boldsymbol{u}'_1 &= \{u[0],u[1],...,u[-1]\}^T\\ &\vdots\\ \boldsymbol{u}'_{N-1} &= \{u[0],u[-N+1],...,u[-1]\}^T.\end{aligned} \tag{3}$$

The aligned subsequences are then superposed sample by sample and averaged, yielding the all-phase sequence

$$\begin{aligned}\boldsymbol{y}_{ap} &= \frac{1}{N}\sum_{i=0}^{N-1}\boldsymbol{u}'_i = \frac{1}{N}\{Nu[0],(N-1)u[1]+u[-N+1],\\ &...,u[N-1]+(N-1)u[-1]\}^T.\end{aligned} \tag{4}$$

As shown in Fig. 2, the segmentation rule in (2) and the reconstruction rule in (4) together perform a weighted recombination of the acquired data, thus completing the preprocessing of the APFFT. Therefore, applying an N-point FFT to $\boldsymbol{y}_{ap}$ is equivalent to performing an FFT on the original (2$N$−1)-point segment with a symmetric triangular weighting, expressed as

$$Y_{AP}(k) = \frac{1}{N}\sum_{n=0}^{N-1} y_{ap}[n]e^{-j2\pi kn/N} = \frac{1}{N}\sum_{n=-(N-1)}^{N-1}\frac{N-|n|}{N}u[n]e^{-j2\pi kn/N} \tag{5}$$

where $k$ is the integer FFT bin index.

To analyze the phase-estimation property, a single-tone signal is considered and only its positive-frequency component is retained as

$$u[n] = \frac{A}{2}e^{j(\frac{2\pi\beta}{N}n+\varphi_0)} \tag{6}$$

where $A$ is the peak amplitude, $\beta$ denotes the true normalized frequency location of the signal in bin units, and $\varphi_0$ is the phase of the center sample $u[0]$.

Substituting (6) into (5), the APFFT spectrum can be expressed as

$$Y_{AP}(k) = \frac{A}{2N^2}\frac{\sin^2[\pi(\beta-k)]}{\sin^2[\pi(\beta-k)/N]}e^{j\varphi_0} \approx \frac{A}{2}\text{sinc}^2(\delta_k)e^{j\varphi_0}. \tag{7}$$

Here, $\delta_k=|\beta-k|$ represents the normalized frequency-bin offset. When $N$ is sufficiently large, (7) can be simplified using the infinitesimal equivalence. Equation (7) further indicates that the APFFT provides an unbiased estimate of the center-sample phase [36].

According to the preceding analysis and the frequency-locking scheme shown in Fig. 1, the REF and DUT signals can be expressed as

$$x_r(t) = A_r\cos(2\pi f_r t+\varphi_r) \tag{8}$$

$$x_d(t) = A_d\cos(2\pi f_d t+\varphi_d). \tag{9}$$

At the $n$-th phase-measurement instant $t_n$, the corresponding phases are

$$\varphi_{r,n} = 2\pi f_r t_n + \varphi_r \tag{10}$$

$$\varphi_{d,n} = 2\pi f_d t_n + \varphi_d. \tag{11}$$

Therefore, the differential phase between the REF and DUT signals is

$$\Delta\varphi_n = \varphi_{r,n} - \varphi_{d,n} = 2\pi(f_r - f_d)t_n + (\varphi_r - \varphi_d). \tag{12}$$

For two adjacent phase-measurement instants, the change in the differential phase is

$$\Delta\varphi_n - \Delta\varphi_{n-1} = 2\pi(f_r - f_d)(t_n - t_{n-1}) \tag{13}$$

The measurement instants corresponding to the $n$-th and ($n$−1)-th APFFT-based estimates are denoted as $t_n$ and $t_{n-1}$, respectively. Let $T_p=t_n-t_{n-1}$ denote the phase-measurement interval. Then, the frequency deviation between the REF and DUT signals can be obtained from the change in their phase difference over $T_p$ as

$$\Delta f = f_r - f_d = \frac{\Delta\varphi_n - \Delta\varphi_{n-1}}{2\pi(t_n - t_{n-1})} = \frac{\Delta\varphi_n - \Delta\varphi_{n-1}}{2\pi T_p}. \tag{14}$$

In practical implementation, the phases extracted by APFFT are wrapped within the principal phase interval [37]. the frequency-deviation estimate is written as

$$\Delta\hat{f} = \frac{\Delta\varphi_n - \Delta\varphi_{n-1} + 2\pi C_n}{2\pi T_p} \tag{15}$$

where $C_n$ is the integer wrap count used to compensate for the 2$\pi$ phase ambiguity between two phase-measurement instants.

Increasing $T_p$ improves the frequency-deviation resolution because a given phase-estimation error is translated into a smaller frequency-deviation error over a longer measurement interval. A counter-assisted unwrapping strategy is used to obtain $C_n$ [26], so that the correct accumulated phase change can be recovered for frequency-deviation extraction.

### *B. Bias Analysis of FFT- and APFFT-Based Phase Estimation*

To clarify the advantage of APFFT in phase estimation, the phase response of the conventional FFT is analyzed under the same single-tone signal model. The normalized $N$-point FFT is directly applied to the $N$-point sequence $\{u[0], u[1],\ldots, u[N-1]\}$. Substituting (6) into the FFT definition gives

$$\begin{aligned} Y_{FFT}[k] &= \frac{A}{2N}\frac{\sin\left[\pi\left(\beta-k\right)\right]}{\sin\left[\pi\left(\beta-k\right)/N\right]}e^{j\left[\varphi_0+\frac{N-1}{N}(\beta-k)\pi\right]} \\ &\approx \frac{A}{2}\sin c(\delta_k)e^{j(\varphi_0+\delta_k\pi)}. \end{aligned} \tag{16}$$

Compared with the APFFT result in (7), the FFT spectrum in (16) contains an additional phase term $\delta_k\pi$, which depends on the frequency-bin offset. Therefore, as shown in Fig. 3, when the signal frequency is not exactly located at an integer FFT bin, the FFT-based phase estimate deviates from the center-sample phase $\varphi_0$. In contrast, the APFFT phase estimate is only determined by $\varphi_0$, indicating that its phase estimate remains unbiased with respect to the center-sample phase. Moreover, the amplitude response of APFFT contains a squared sinc-shaped factor, whereas that of the conventional FFT contains a single sinc-shaped factor. This squared response provides stronger sidelobe attenuation and better spectral concentration around the dominant frequency component, thereby improving spectral-leakage suppression. Therefore, APFFT is more suitable for high-precision phase extraction in the proposed frequency-locking method.

### *C. Error Analysis*

For a finite-length single-tone signal, the FFT spectrum shows a sinc-like response, with its main-lobe center corresponding to the true frequency. Accordingly, the peak spectral bin $k^*$ obtained by APFFT is selected as the nearest integer-bin estimate of the true frequency-bin index $\beta$, and is subsequently used for frequency-deviation estimation.

$$Y(k^*) = S(k^*) + \eta_{ap}(k^*) \tag{17}$$

where $S(k^*)$ denotes the useful signal component, and $\eta_{ap}(k^*)$ denotes the equivalent noise component. According to the APFFT model described above, $S(k^*)$ can be expressed as

$$S(k^*) = \frac{A}{2}\sin c^2(\delta_{k^*})e^{j\varphi_0}. \tag{18}$$

The all-phase preprocessing can be interpreted as an equivalent triangular weighting operation, as shown in Fig. 2. If the input noise variance is $\sigma_n^2$, then the noise power at the APFFT output is

$$E\left[\left|\eta_{ap}(k)\right|^2\right] \approx \frac{2\sigma_n^2}{3N}. \tag{19}$$

Under high-SNR conditions, the phase error can be obtained by projecting the noise phasor onto the quadrature direction of the useful signal phasor as

$$\varepsilon_\varphi(k^*) = \frac{\mathrm{Im}[\eta_{ap}(k^*)e^{-j\varphi_0}]}{\left|S(k^*)\right|}. \tag{20}$$

Thus, the variance of the phase error is obtained as

$$\mathrm{var}[\varepsilon_\varphi(k^*)] = \frac{4\sigma_n^2}{3NA^2\sin c^4(\delta_{k^*})}. \tag{21}$$

The expression in (21) is consistent with that reported in [38]. By combining (14) and (21), the variance of the estimated frequency deviation can be approximated as

$$\mathrm{var}(\Delta f) = \frac{4\,\mathrm{var}[\varepsilon_\varphi(k^*)]}{4\pi^2T_p^2} = \frac{4\sigma_n^2}{3\pi^2NA^2T_p^2\sin c^4(\delta_{k^*})}. \tag{22}$$

**1) *System Thermal Noise***

The system thermal noise is modeled as an additive white Gaussian noise (AWGN) process with zero mean and variance $\sigma_w^2$. For two channels with equal signal amplitudes and noise variances, the SNR is defined as $SNR = A^2/2\sigma_w^2$. The frequency measurement precision is characterized by the standard deviation (STD) of the estimated frequency deviation. Accordingly, the contribution of the system thermal noise to the measurement precision is expressed as

$$\mathrm{std}_{th}(\Delta f) = \frac{\sqrt{2}}{\pi T_p\sqrt{3N\cdot SNR}\sin c^2(\delta_{k^*})}. \tag{23}$$

**2) *ADC Quantization Noise***

The ADC quantization process introduces an amplitude discretization error into the sampled sequence. The quantization error can be modeled as an additive white noise process, $e_q[n]$, uniformly distributed over [$-\Delta/2$, $\Delta/2$], where $\Delta$ is the quantization step size. For an ADC with a full-scale range of $2A_{FS}$ and a resolution of $b$ bits, the step size is $\Delta = 2A_{FS}/2^b$. The variance of the quantization noise $\sigma_q^2$ is given by

$$\sigma_q^2 = \int_{-\Delta/2}^{\Delta/2}x^2\frac{1}{\Delta}dx = \frac{\Delta^2}{12} = \frac{A_{FS}^2}{3\cdot 2^{2b}}. \tag{24}$$

Assuming that the signal amplitude $A$ fully utilizes the ADC full-scale range of the ADC, the standard deviation of the frequency-measurement error due to quantization is

$$std_q(\Delta f) = \frac{1}{3\pi T_p\sqrt{N}2^{b-1}\sin c^2(\delta_{k^*})}. \tag{25}$$

**3) *Clock* and *Sampling Jitter***

Sampling clock jitter introduces timing errors that translate into amplitude errors in the sampled sequence. Let the ideal sampling time be $t_m=mT_S$. In the presence of jitter, the actual sampling occurs at $t_m+\tau_j[n]$, where $\tau_j[n]$ represents a zero-mean random timing error with variance $\sigma_t^2$. For the input sinusoidal signal, the jitter-induced amplitude error can be approximated using a first-order Taylor expansion as

$$e_j[n] \approx \left.\frac{du(t)}{dt}\right|_{t=t_m}\cdot\tau_j[n] = 2\pi fA\cos\left(2\pi ft_n+\varphi\right)\tau_j[n]. \tag{26}$$

Unlike thermal noise and quantization noise, the jitter-induced error is signal-dependent because it is proportional to the instantaneous slope of the input signal. For the selected spectral bin $k^*$, the jitter-induced perturbation at the APFFT output can be written as

$$\eta_j(k^*) = \frac{1}{N}\sum_{n=-(N-1)}^{N-1}\frac{N-|n|}{N}e_j[n]e^{-j2\pi k^*n/N}. \tag{27}$$

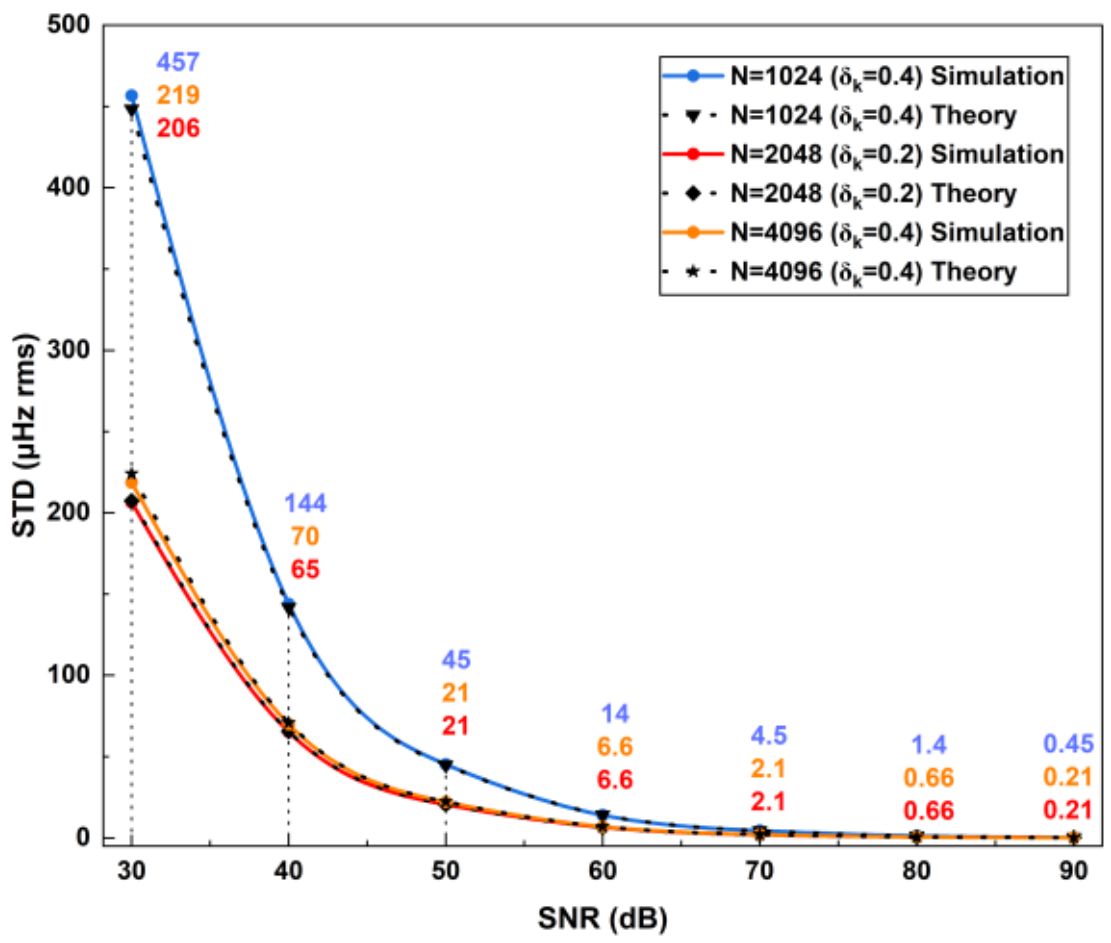

Fig. 4. Simulated frequency-deviation measurement error due to system thermal noise.

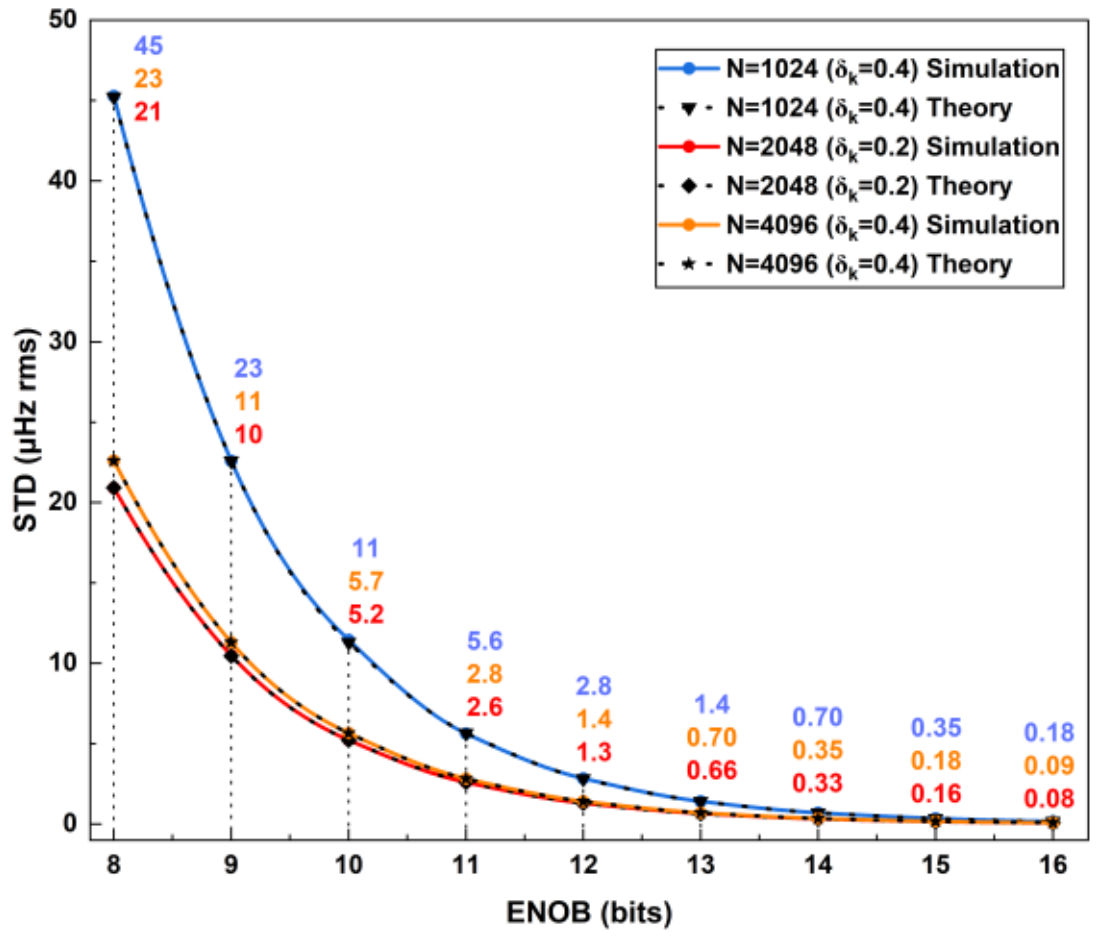

Fig. 5. Simulated frequency-deviation measurement error due to ADC quantization noise.

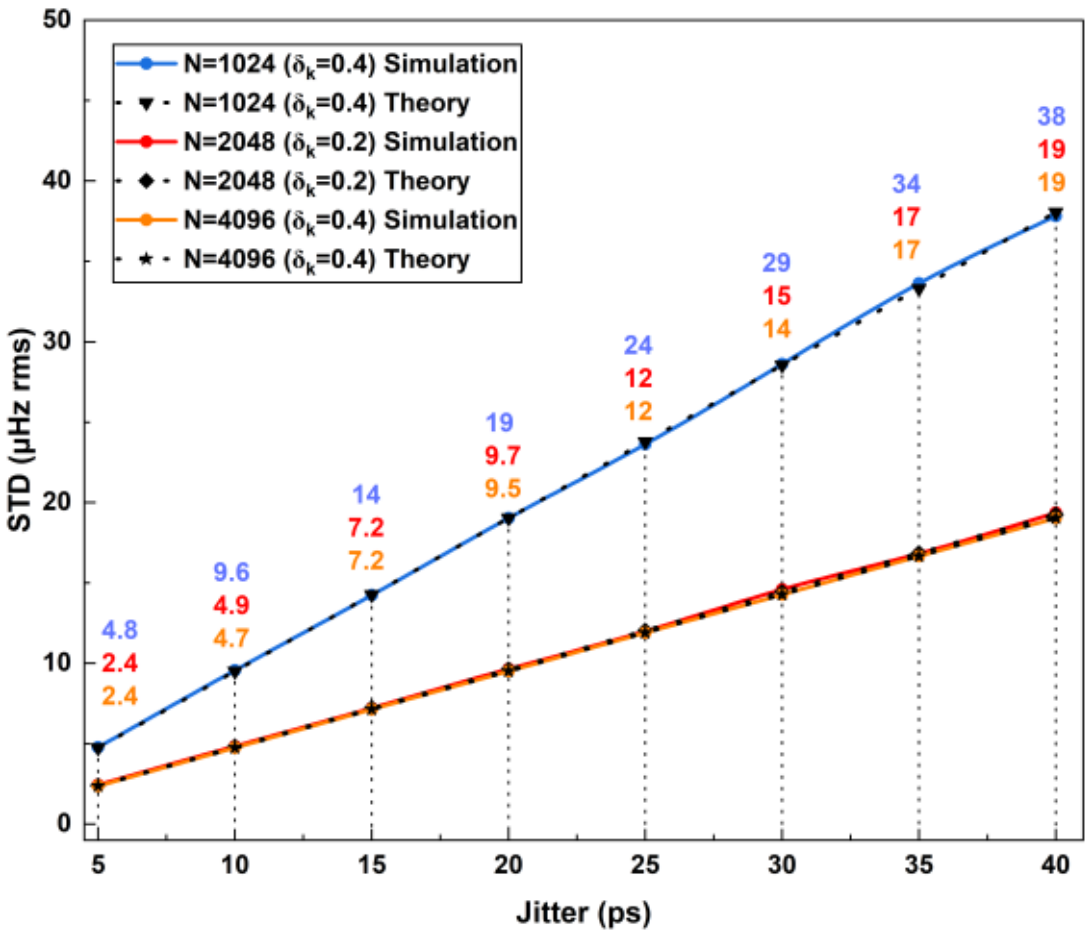

Fig. 6. Simulated frequency-deviation measurement error due to clock and sampling jitter.

Under high-SNR conditions, the phase error is obtained by projecting $\eta_j(k^*)$ onto the quadrature direction of the useful APFFT phasor. Averaging over the signal phase gives

$$\mathrm{var}[\varepsilon_j(k^*)] = \frac{(2\pi f\sigma_t)^2}{N^2 \operatorname{sinc}^4(\delta_{k^*})} \sum_{n=-(N-1)}^{N-1} \left(\frac{N-|n|}{N}\right)^2 \left[\cos^2\left(\frac{2\pi\delta_{k^*} n}{N}\right) + \frac{1}{2}\right]. \quad (28)$$

Thus, the STD of the frequency-measurement error caused by sampling clock jitter is

$$std_j(\Delta f) = \frac{2f\sigma_t}{T_p\sqrt{N}\operatorname{sinc}^2(\delta_{k^*})} \sqrt{\frac{2}{3} + \frac{2[4\pi\delta_{k^*} - \sin(4\pi\delta_{k^*})]}{(4\pi\delta_{k^*})^3}}. \quad (29)$$

## III. Simulation

To verify the theoretical error model derived in Section II-C, Monte Carlo simulations were performed on MATLAB platform. The sampling frequency was set to $F_s$ = 100 MHz, and the input signal frequency was $f$ = 10 MHz. The phase-measurement interval was set to $T_p$ = 1 s. Three APFFT lengths were considered, namely $N$ = 1024, $N$ = 2048, and $N$ = 4096. Since the true normalized frequency location is $\beta = Nf / F_s$, the corresponding normalized frequency-bin offsets were $\delta_k$ = 0.4, 0.2, and 0.4. For each condition, the frequency deviation was estimated using the APFFT-based phase extraction and phase-difference method described in Section II. For each parameter set, 10,000 Monte Carlo trials were performed to obtain the statistical characteristics of the frequency-deviation measurement error. The standard deviation of the estimated frequency deviation was then calculated and compared with the theoretical prediction.

Figs. 4-6 show the simulated and theoretical frequency-deviation STD caused by system thermal noise, ADC quantization noise, and sampling clock jitter, respectively. In Fig. 4, additive white Gaussian noise was added to the sampled sequences, and the SNR was swept from 30 dB to 90 dB. In Fig. 5, the ADC effective number of bits (ENOB) was swept from 8 bits to 16 bits to evaluate the quantization-noise contribution. In Fig. 6, the RMS sampling-clock jitter was swept from 5 ps to 40 ps. The simulation results show good agreement with the theoretical predictions of (23), (25), and (29) under all three noise conditions, confirming the validity of the error models derived in Section II-C.

The simulation results also indicate that selecting the APFFT length $N$ requires a tradeoff rather than simply using the largest value. Although increasing N generally enhances noise suppression by involving more samples in the all-phase preprocessing and FFT operation, the estimation precision is also influenced by the frequency-bin offset $\delta_k$. As a result, a larger N does not always lead to a proportional improvement. In this simulation, $N$ = 1024 provides insufficient precision, while $N$ = 2048 with $\delta_k$ = 0.2 achieves performance close to, and in some cases slightly better than, $N$ = 4096 with $\delta_k$ = 0.4. Therefore, $N$ = 2048 offers a favorable tradeoff among frequency-deviation precision, FPGA memory usage, FFT latency, and computational resource consumption, and is selected for hardware implementation.

## IV. Implementation and Results

A hardware prototype was developed to validate the proposed APFFT-based frequency-locking scheme. In this prototype, a VCO was adopted as the DUT, and a rubidium atomic clock was used as the high-stability reference source. Based on this experimental platform, the effectiveness of the proposed method was evaluated through frequency-deviation measurements and closed-loop frequency-locking demonstrations.

TABLE I
COMPARISON OF REPRESENTATIVE FREQUENCY-DEVIATION EXTRACTION METHODS FOR FREQUENCY LOCKING

| Ref. | Method | Features | On-board Implementation | Accuracy |
|---|---|---|---|---|
| [25] | DQD | Time-domain single IQ demodulation, NCO-sensitive | Moderate | Medium |
| [26] | ADC-based-DMTD | Time-domain differential demodulation, NCO-noise-suppressed | Complex | Medium-high |
| [29] | FFT | Frequency-domain spectral phase extraction, leakage-affected | Easy | Limited |
| This work | APFFT | Frequency-domain all-phase phase extraction, leakage-suppressed | Moderate | High |

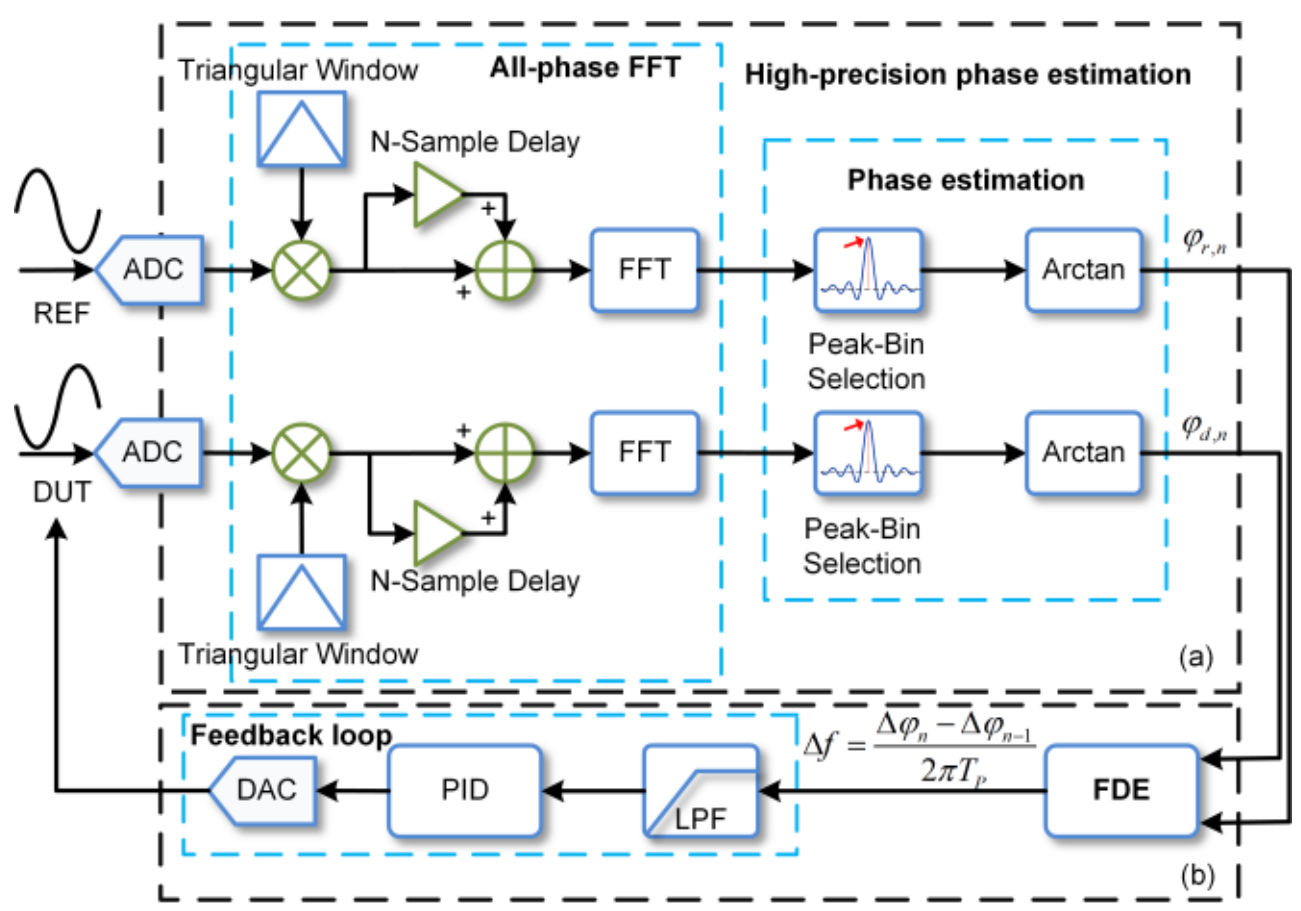

Fig. 7. Hardware architecture of the proposed APFFT-based frequency-locking prototype. (a) Dual-channel APFFT-based phase-estimation module for the REF and DUT signals. (b) Frequency-deviation extraction and digital feedback loop for frequency locking.

### A. Hardware Implementation

Fig. 7 presents the block diagram of the implemented hardware system. In the experimental setup, an AOCJY1 10-MHz VCO was used as the DUT, while a SAFRAN LPFRS-01 10-MHz rubidium atomic clock was used as the high-stability REF. Motivated by the error analysis, the REF and DUT signals were synchronously acquired by a 16-bit ADS5263 ADC operating at 100 MS/s to reduce quantization-induced error. An LMK04610 jitter cleaner provided low-jitter sampling and system clocks with a typical rms jitter of 65 fs, to mitigate clock-jitter-induced frequency-deviation error. The sampled data were then processed in real time using a Xilinx XC7K325TFFG900-2 FPGA, where dual-channel APFFT-based phase estimation, frequency-deviation extraction, and digital feedback control were implemented.

As shown in Fig. 7(a), after synchronous sampling, the REF and DUT signals are processed in parallel by two identical APFFT-based channels to obtain unbiased phase estimates. In the FPGA implementation, each APFFT-based channel was realized as an independent processing pipeline. The synchronously sampled data were first buffered to form a (2N-1)-point segment, which was then processed by the all-phase preprocessing module. In this module, triangular-window weighting and folding accumulation were performed to reconstruct an N-point all-phase sequence. The reconstructed sequence was sent to the FFT core, followed by peak-bin selection and phase calculation from the real and imaginary spectral components using an arctangent unit. The dual-channel FPGA architecture enables parallel and consistent phase extraction for the REF and DUT signals, while the APFFT structure reduces the sensitivity of the phase estimation to spectral leakage.

TABLE II
FPGA RESOURCE UTILIZATION OF THE APFFT-BASED FREQUENCY-LOCKING METHOD IMPLEMENTED ON AN XC7K325TFFG900-2

| FPGA resources | Occupied | Available | Utilization |
|---|---|---|---|
| LUT | 39655 | 203800 | 19.46% |
| LUT RAM | 3572 | 64000 | 5.58% |
| FF | 52177 | 407600 | 12.80% |
| BRAM | 102 | 445 | 22.92% |
| DSPs | 70 | 840 | 8.33% |
| BUFG | 18 | 32 | 56.25% |
| MMCM | 4 | 10 | 40.00% |

As shown in Fig. 7(b), the extracted phase estimates were delivered to the FDE block, where the instantaneous frequency deviation was calculated from the phase evolution of the two channels over the phase-measurement intervals. The frequency-deviation signal was further processed by a low-pass filter (LPF) and digital proportional–integral–derivative (PID) controller. The resulting control word was converted by the digital-to-analog converter (DAC) into the tuning voltage for the DUT, thereby disciplining the VCO to the rubidium-clock reference.

To clarify the advantages and limitations of the proposed method relative to representative frequency-error extraction strategies used in digital frequency-locking systems, a qualitative comparison is provided in Table I. And the Table II summarizes the FPGA resource utilization of the implemented design on the Xilinx XC7K325TFFG900-2 device. The consumption of logic, memory, and arithmetic resources remained at moderate levels. The current design operates at a 100MHz system clock. For the $N$ = 2048 APFFT implementation, the initial data buffering stage requires 4096 sampling-clock cycles, corresponding to 40.96 µs. The all-phase preprocessing introduces an additional latency of 0.03 µs. The FFT IP core has a latency of 42.56 µs from the first input sample to the first output sample. The peak-bin selection module and the CORDIC module introduce latencies of 10.27 µs and 0.37 µs. Therefore, the total latency from the sampling trigger to the first valid phase-estimation result is 94.19 µs.

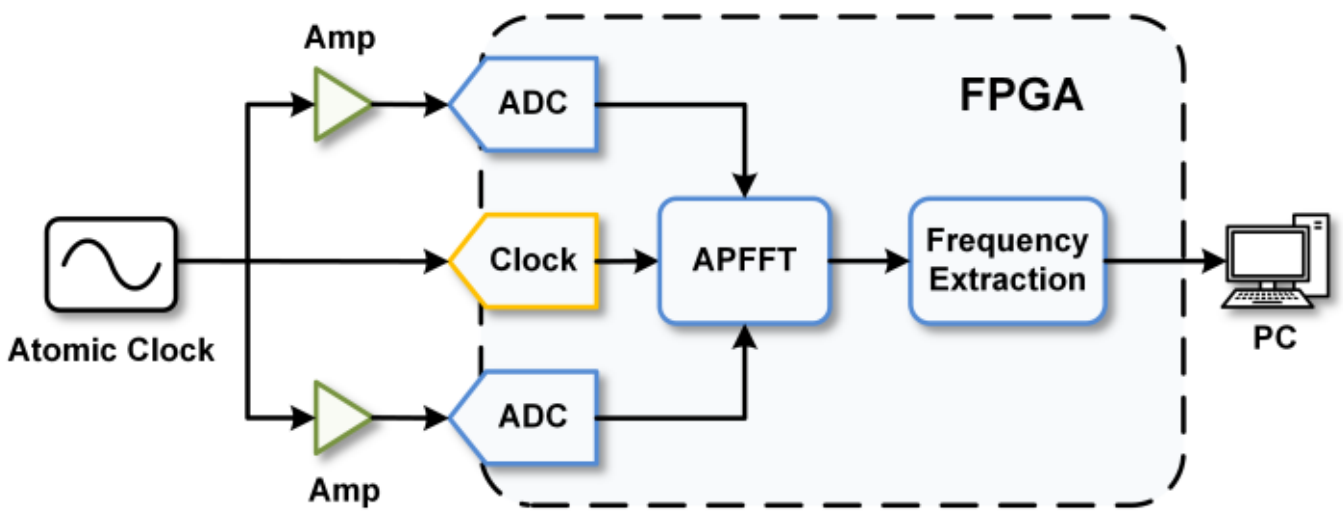

Fig. 8. Block diagram of frequency-measurement noise-floor test.

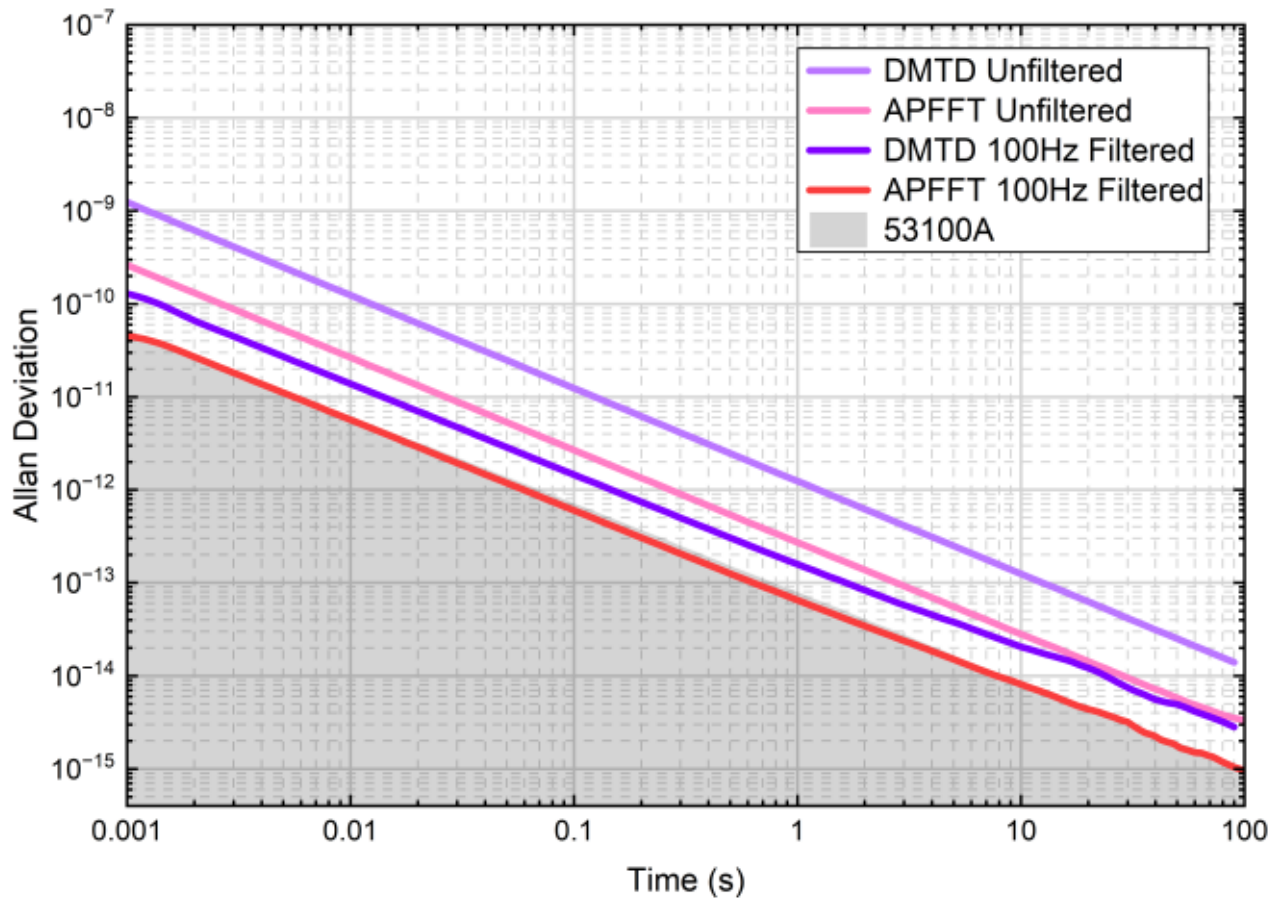

Fig. 9. Allan deviation of the frequency-deviation measurement using the APFFT ($N$=2048) and ADC-based-DMTD methods.

After the processing pipeline is filled, the system can update the phase-estimation result every 20.48 μs.

*B. Noise Floor of Frequency Measurement*

The intrinsic frequency-measurement noise floor of the implemented system was first evaluated using a common-source test [22], as shown in Fig. 8. A 10-MHz rubidium atomic clock SAFRAN LPFRS-01 was split and synchronously digitized by two ADC channels. The sampled data from the REF and DUT channels were processed on the FPGA using the proposed APFFT-based method, and the extracted frequency-deviation results were transmitted to the host computer for statistical analysis. Because both channels were driven by the same source, the ideal frequency deviation between them was zero. Thus, the residual fluctuation measured in this configuration mainly reflects the intrinsic noise floor of the frequency-measurement system.

In this test, the APFFT length was set to $N$ = 2048, corresponding to a frequency-bin offset of $\delta_k = 0.2$ for $F_s = 100$ MHz and $f = 10$ MHz. A 1-s measurement interval was used for the STD-based noise-floor evaluation. The measured system SNR was 72.05 dB, including both acquisition-chain thermal noise and ADC quantization noise, which corresponds to an equivalent ENOB of 11.68 bits. This value was used as the input parameter for the theoretical noise-floor prediction. The contribution of sampling-clock jitter was much smaller than that of the acquisition-chain noise and was therefore neglected. Under these conditions, the predicted frequency-measurement noise floor was approximately 1.64 μHz rms, whereas the experimentally measured value was 2.31 μHz rms. The slightly higher measured noise floor can be attributed to residual hardware nonidealities not fully included in the theoretical model. Nevertheless, the experimental result agrees well with the theoretical prediction, validating the proposed noise-floor model.

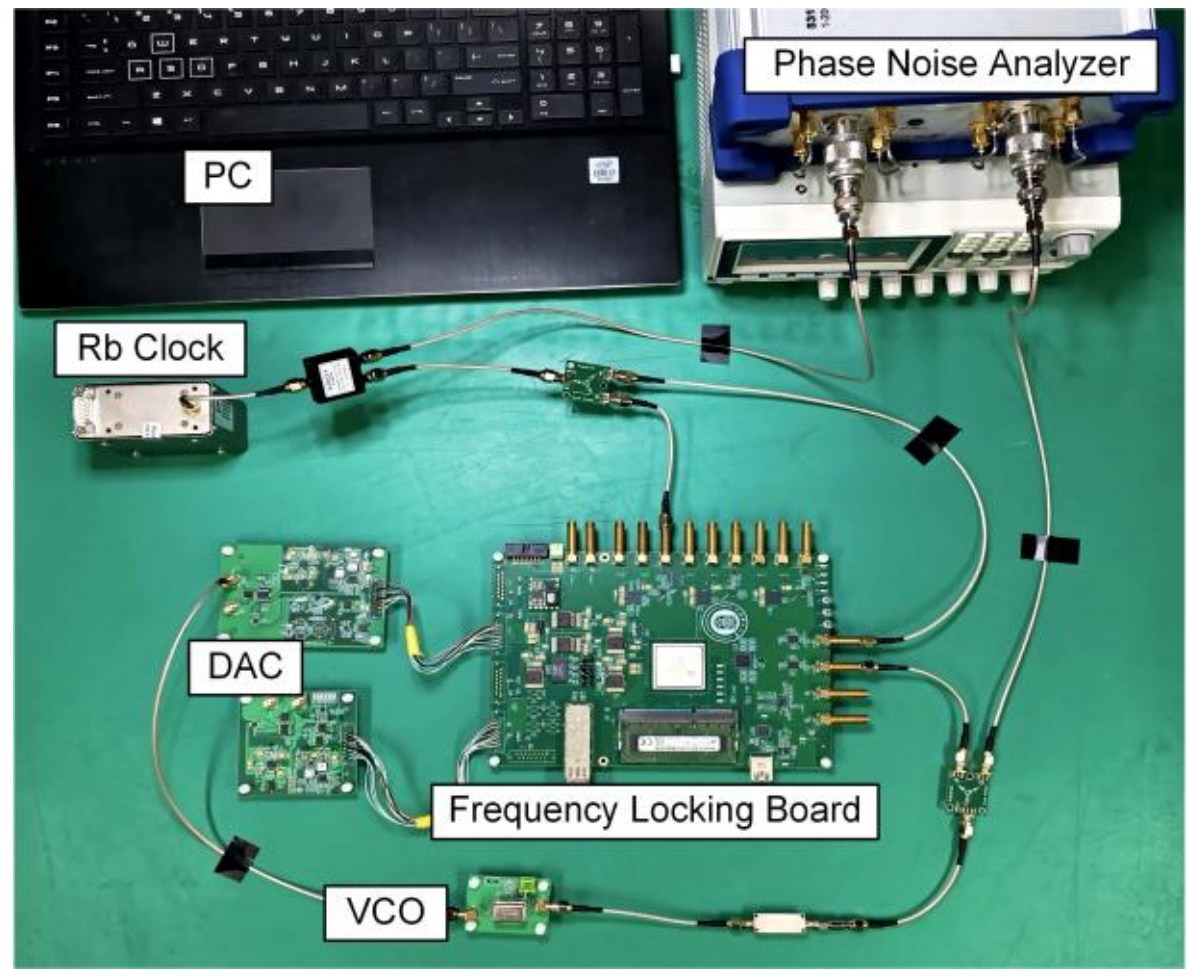

Fig. 10. Scene of the VCO frequency locking experiment.

Fig. 9 shows the Allan deviation of the frequency-deviation measurements obtained using the APFFT-based and ADC-based-DMTD [26] methods on the same test platform. The frequency-deviation data were recorded at an update rate of 1 kHz over a total duration of 400 s. Both the unfiltered results and the results after applying a 100-Hz digital low-pass filter are presented. For comparison, the gray region indicates the background noise floor of the 53100A phase noise analyzer under a 500-Hz filter bandwidth and a 1-kHz sampling rate. Over the measured averaging-time range, the APFFT-based method consistently outperforms the ADC-based-DMTD method, as evidenced by its lower Allan deviation. With the additional 100-Hz filtering, the Allan deviation is further reduced because the residual high-frequency measurement noise is suppressed. The APFFT-based method achieves Allan deviations of approximately $6.51 \times 10^{-14}$ and $8.10 \times 10^{-15}$ at averaging times of 1 s and 10 s, respectively, which are about one order of magnitude lower than the corresponding values of $1.58 \times 10^{-13}$ and $2.05 \times 10^{-14}$ obtained using the ADC-based-DMTD method.

*C. Performance of Frequency-Locking*

As shown in Fig. 10, a frequency-locking experimental platform was built to verify the proposed APFFT-based locking method. The platform included an AOCJY1 10-MHz VCO as the DUT, a SAFRAN LPFRS-01 10-MHz rubidium atomic clock as the high-stability reference, the implemented frequency-locking board, and a DAC that generated the tuning voltage for the VCO. The rubidium atomic clock provides a nominal frequency stability of $2 \times 10^{-11}$ at 1 s. A 53100A phase noise analyzer was also used as an external reference instrument to independently evaluate the VCO stability before and after locking.

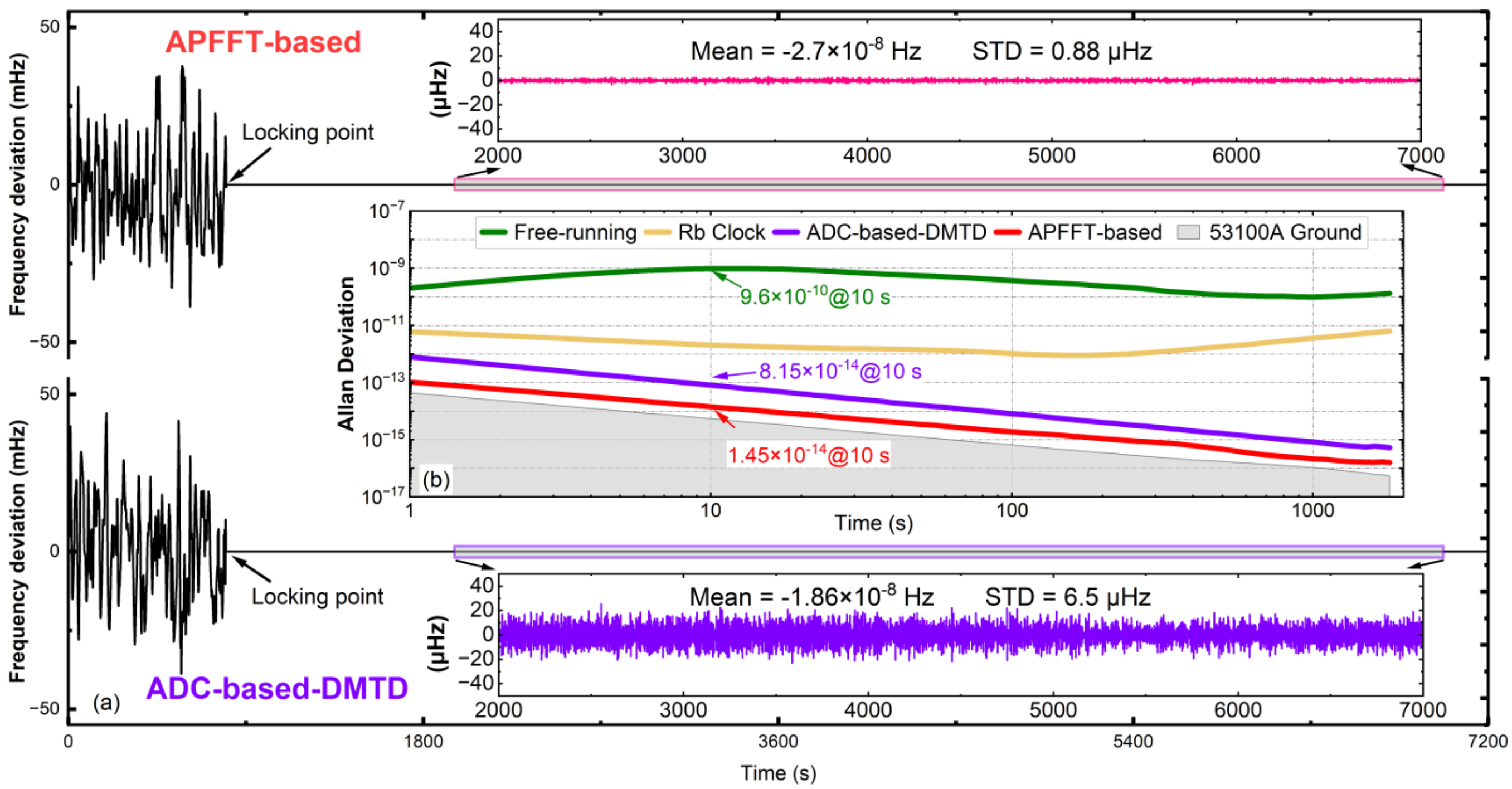

Fig. 11. Frequency-locking results of the VCO using the proposed APFFT-based method and the ADC-based-DMTD method. (a) Frequency-deviation traces before and after locking, together with zoomed-in views of the locked-state fluctuations measured by the 53100A phase noise analyzer. (b) Allan deviation of the free-running VCO and the locked results obtained by the two methods. The gray region indicates the noise floor of the 53100A phase noise analyzer.

The VCO performance after locking was first characterized using the self-developed APFFT-based frequency-locking electronics. The frequency-deviation estimates obtained by the APFFT-based module were transmitted to the host computer for statistical analysis. With a 1-s measurement period, the standard deviation of the locked frequency deviation was 4.04 μHz rms, slightly higher than the intrinsic electronics noise floor of 2.31μHz rms. This measured value includes the intrinsic noise floor of the APFFT-based module, the residual frequency fluctuation of the locked VCO, and the frequency fluctuation of the rubidium-clock reference.

To independently verify the locking performance, the VCO output was further characterized using a 53100A phase noise analyzer. The analyzer was operated with a 1-s gate time, and its built-in filter bandwidth was limited to 5 Hz. Under this configuration, the STD-based frequency-deviation noise floor of the 53100A was approximately 0.35 μHz, which is lower than the residual fluctuations measured in the locking experiments and therefore sufficient for evaluating the locked-state performance. Fig. 11(a) shows the frequency-deviation traces of the VCO before and after locking. In the free-running state, the VCO exhibits frequency fluctuations at the millihertz level. Once the feedback loop is enabled, the frequency deviation is rapidly suppressed and enters the locked state. Both the proposed APFFT-based method and the previously reported ADC-based-DMTD method are able to realize frequency locking under the same experimental conditions. However, the zoomed-in traces clearly show that the APFFT-based method produces a much smaller residual fluctuation after locking. The residual frequency-deviation STD measured by the 53100A is 0.88 μHz rms or the APFFT-based method, whereas it is 6.5 μHz rms for the ADC-based-DMTD method. This result demonstrates that the proposed method reduces the locked-state frequency fluctuation by more than a factor of about seven compared with the ADC-based-DMTD method.

The improvement is further confirmed by the Allan-deviation results shown in Fig. 11(b). The free-running VCO exhibits an Allan deviation of $9.6 \times 10^{-10}$ at 10 s. After locking, the Allan deviation is significantly reduced for both methods, indicating the effectiveness of the digital frequency-locking scheme. Nevertheless, the APFFT-based method provides a markedly better stability improvement. At an averaging time of 10 s, the Allan deviation is reduced to $8.15 \times 10^{-14}$ at 10 s using the ADC-based-DMTD method, while the proposed APFFT-based method further reduces it to $1.45 \times 10^{-14}$ at 10s. The APFFT-based result is also closer to the noise floor of the 53100A, as indicated by the gray shaded region, and approaches the stability level represented by the rubidium-clock reference comparison.

These experimental results show that the proposed APFFT-based frequency-deviation extraction method provides a lower measurement noise contribution in the feedback loop, leading to a smaller residual frequency deviation and better frequency stability. Compared with the ADC-based-DMTD method, the proposed method enables more effective frequency locking and demonstrates its advantage for high-precision digital frequency-control applications.

## V. CONCLUSION

This article has presented an APFFT-based digital frequency-locking method for high-precision frequency control. The core of the proposed method is to extract the frequency deviation from unbiased frequency-domain phase estimates obtained by two parallel APFFT-based channels. Compared with conventional digital frequency-error extraction methods, the proposed method makes more effective use of the phase information contained in the sampled signals and provides a lower measurement-noise contribution to the feedback loop.

The bias and noise characteristics of the FFT- and APFFT-based phase-estimation methods were analyzed, and an error model was established to guide the selection of key parameters such as the APFFT length and the measurement interval. The proposed method was implemented on an FPGA using a dual-channel APFFT processing architecture, including all-phase preprocessing, FFT operation, peak-bin selection, and phase extraction.

The effectiveness of the proposed method was further verified in a VCO frequency-locking experiment. Independent measurements using a 53100A phase noise analyzer showed that the residual frequency-deviation STD after locking was 0.88 μHz rms for the APFFT-based method, compared with 6.5 μHz rms for the previously reported ADC-based-DMTD method under the same experimental conditions. The Allan deviation was also significantly improved; at an averaging time of 10 s, the APFFT-based method achieved $1.45 \times 10^{-14}$, outperforming the ADC-based-DMTD result of $8.15 \times 10^{-14}$.

These results demonstrate that the APFFT-based method is not only a high-precision frequency-deviation extraction technique, but also an effective digital locking approach for compact and high-stability frequency-control systems. Owing to its fully digital architecture, low residual measurement noise, and FPGA implementability, the method is promising for oscillator disciplining, pulse-laser repetition-rate locking, dual-comb and THz-TDR systems, time-frequency synchronization, and other precision instrumentation applications requiring high stability and compact implementation.

## ACKNOWLEDGMENT

The authors would like to thank Jiaqi Zhou from the Shanghai Institute of Optics and Fine Mechanics, Chinese Academy of Sciences, and Qing Sun at the National Institute of Metrology for useful discussions. The authors also extend their gratitude to Qi Shen from the University of Science and Technology of China for his insightful discussions.